\documentclass[%preprint,
superscriptaddress,
 amsmath,amssymb,
 aps,
reprint
]{revtex4-2}

\usepackage{color}
\usepackage{graphicx}% Include figure files
\usepackage{dcolumn}% Align table columns on decimal point
\usepackage{bm}% bold math
\usepackage{epstopdf}
\usepackage{gensymb}
\usepackage{hyperref}
\usepackage{float}

\begin{document}

%\preprint{APS/123-QED}

\title{Analytical model for the remote epitaxial potential}

\author{Jason K. Kawasaki} \email{Author to whom correspondence should be addressed: Jason Kawasaki, jkawasaki@wisc.edu}
\affiliation{Materials Science and Engineering, University of Wisconsin-Madison, Madison, WI 53706, United States of America}

\author{Quinn T. Campbell}
\affiliation{Sandia National Laboratories, Albuquerque, NM 87123, United States of America}

\date{\today}% It is always \today, today,

\begin{abstract}
We propose an analytical model for the remote bonding potential of the substrate that permeates through graphene during remote epitaxy. Our model, based on a Morse interatomic potential, includes the attenuation due to the increased film-substrate separation and due to graphene free carrier screening. Compared with previous slab density functional theory calculations, which use the electrostatic potential as a proxy for bonding, our analytical model includes covalent and van der Waals bonding interactions, includes screening (which is often ignored), and is based on simple, physically interpretable, and well benchmarked parameters that build understanding.
We show that for typical graphene free carrier concentrations of order $10^{12}$ cm$^{-2}$, the magnitude of $|\phi_{remote}|$ for most semiconductor and oxide substrates is few to tens of meV, similar to the van der Waals potential of graphene. This suggests interference between the graphene and remote substrate potentials must be considered when interpreting experiments on remote epitaxy. Furthermore, we show that (1) the strength of the remote potential is tunable by screening (graphene carrier density), (2) the remote potential is vanishingly weak through two or more graphene layers, (3) the spatial extent of the potential, rather than degree of ionicity, controls the strength of the permeated bonding potential. Our model provides a simple framework for benchmarking direct measurements of the remote epitaxial potential, and we propose several experimental paths to measure this quantity. De-convolving the effects of the native remote potential, graphene defects, and tunable growth kinetics is key to understanding and tuning the mechanisms of remote epitaxy.

\end{abstract}

\maketitle

Bonding is typically considered a very short-range interaction ($\lessapprox$ 3 \AA). Atomic and molecular orbitals (wavefunctions) decay exponentially with distance, and thus nearest neighbor interactions tend to dominate. Remote epitaxy challenges this paradigm. Here it is argued that the lattice potential of a crystalline substrate can permeate through graphene with sufficient strength to template epitaxy of a film on top of a graphene-covered substrate \cite{kim2017remote, kong2018polarity}. In this picture, the graphene is nearly transparent \cite{rafiee2012wetting}: the film adopts epitaxial registry to the buried substrate rather than graphene, and the lattice potential of the graphene itself is assumed to contribute minimally. The graphene spacer merely attenuates the film-substrate interaction strength. This concept is quite useful: due to the weaker film-substrate interactions, films grown on graphene can be exfoliated to produce flexible, stackable, and strain engineered \cite{laduca2024cold, du2021epitaxy} membranes. The weakened interactions may also enable lattice mismatched heteroepitaxy of films with decreased dislocation density due to the ``slippery'' graphene interface. 

The current theoretical understanding of remote epitaxy is based on the density functional theory (DFT)-computed electrostatic potential decay with distance \cite{kong2018polarity, dai2022highly, qu2022long, jiang2019carrier}, which typically ignore interfacial reconstructions and assumes all the atoms are near their bulk like positions (presumably to minimize computational cost). These calculations suggest that substrates with greater bond polarity produce larger electrostatic fluctuations above graphene, and thus can support remote epitaxy through greater number of graphene layers \cite{kong2018polarity}. 

%\textcolor{red}{[alt intro] However, there are several challenges to this understanding. First, there is a need for analytical models built from physically-interpretable parameters, to build a physical understanding beyond DFT-based approaches that can be a black box. Second, current DFT approaches focus on the permeated electrostatic potential through graphene. Electrostatics define important electronic quantities like the workfunction and band alignments, yet is is not clear that electrostatics should drive the formation and structure of crystals. Typically, covalent bond hybridization an van der Waals interactions are more relevant at nearest and next nearest neighbor length scales. Third, DFT treatments often ignore free carrier screening from graphene, ignore the lattice potential of graphene, and ignore possible reconstructions. Finally, comparison to experiments are challenging because most experiments}

However, there are several challenges to this understanding. First, DFT slab calculations can be a computationally expensive black box, making it difficult to interpret the contributions from various terms without additional tests. A simple analytical model is needed to build a physical understanding of the various terms that contribute to the strength of the remote potential, and enable rapid surveys over candidate materials. Second, although electrostatics define important electronic quantities like the workfunction and band alignments, the formation and structure of crystals are instead typically driven by covalent and van der Waals (vdW) bonding interactions between the substrate and film. New models need to consider these covalent and vdW interactions, rather than the pure electrostatics focus of existing DFT, which often does not incorporate any knowledge of the film.

Finally, theories of the remote potential are difficult to benchmark because there are few direct measurements of the remote potential \cite{hibino2009dependence, filleter2008local}. Instead, its strength is typically inferred from the final outcomes of growth experiments \cite{kong2018polarity,kim2017remote,dai2022highly,jiang2019carrier}, which convolve the remote potential with defects and growth kinetics \cite{laduca2025transparent}. Defects and kinetics have a significant impact on growth mechanisms: previous experiments \cite{manzo2022pinhole} demonstrate that since typical adatom diffusion lengths on graphene are 100 nm to microns at typical growth temperatures \cite{liu2011fe, lim2022selective, manzo2022pinhole, liu2013slow}, pinhole-seeded epitaxy, rather than remote epitaxy, dominates if the defect density on graphene exceeds 1 $\mu m^{-2}$ \cite{manzo2022pinhole,laduca2025transparent}. Since most experiments do not report the graphene defect density after the critical anneal step that produces pinholes \cite{manzo2022pinhole}, most experiments to date cannot distinguish remote epitaxy from pinhole epitaxy. An analytical theory of the remote potential, with physically interpretable parameters that can be benchmarked against direct measurements of the remote potential, is required to de-convolve the various factors that dictate the growth mechanisms on graphene-covered surfaces.

Here we develop an analytical model for the remote bonding potential permeation through graphene. Compared to previous DFT-based electrostatic potential calculations, our model (1) is based on simple physical parameters to aid physical understanding rather than black box approaches, (2) includes more physically motivated covalent bonding and vdW interactions between the substrate and film rather than purely electrotatics \cite{mirzanejad2025derivation}, and (3) includes free carrier screening of graphene, which is often ignored \cite{dai2022highly, kim2017remote, jiang2019carrier}. %(4) and avoids the use of adjustable scale factor, complicated background subtraction, and cropping \cite{kong2018polarity}.
We show that graphene is not as transparent as often assumed: for typical graphene carrier densities, the remote potential of the substrate through graphene has an order of magnitude few meV, similar to the the vdW interaction strength of graphene itself. Our model predicts that the spatial extent of the bond potential, rather than bond ionicity, is the primary parameter that controls the strength of the remote potential. Finally, our model identifies graphene screening as an attractive parameter for continuously tuning the strength of the remote potential. We propose direct measurements of the remote potential based on scanning probe and atom scattering techniques to benchmark our model, as a function of gate-tuned graphene carrier density.

\begin{figure}
    \centering
    \includegraphics[width=1\linewidth]{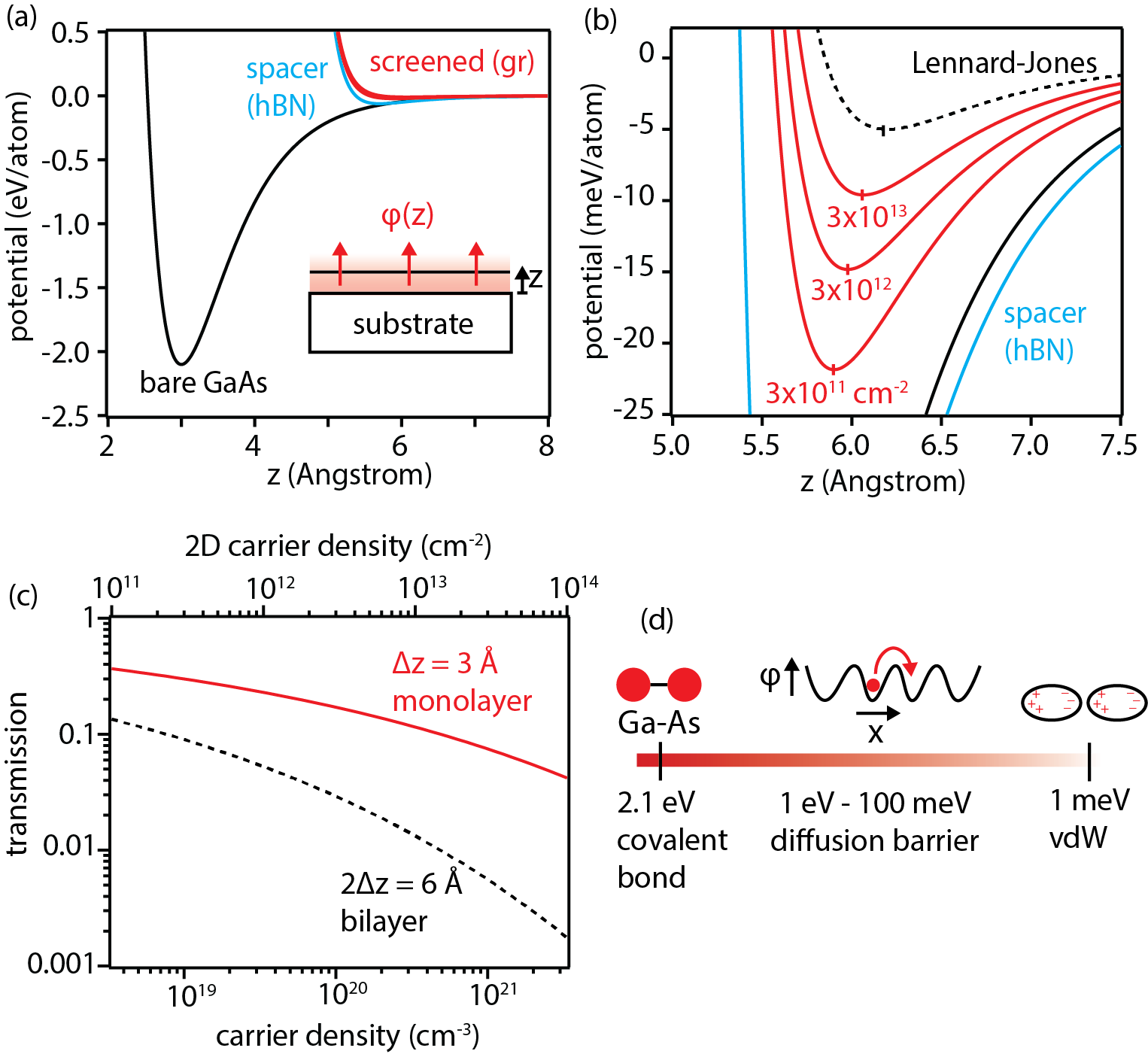}
    \caption{Analytical model for the remote bonding potential through graphene. (a) Morse potential $\phi_{sub}$ for bare GaAs (black), insulating hBN on GaAs (blue, $\phi_{sub}+T_s\phi_{LJ}$), and the screened total potential through monolayer graphene on GaAs (red, $T_s \phi_{sub}+\phi_{LJ}$). $z=0$ corresponds to the surface of the substrate. (b) Zoom in of (a) featuring the screened total potentials for carrier densities $3\times 10^{11}$, $3\times 10^{12}$, and $3\times 10^{13}$ cm$^{-2}$ (red) and the graphene Lennard-Jones potential (dotted black). (c) Free carrier screening in the Thomas-Fermi approximation. (d) Energy scales for covalent bonding, surface diffusion, and van der Waals interactions.}
    \label{fig:screening}
\end{figure}

%Our model initially treats graphene in the ideal limit as a homogeneous electron gas. The key elements are (1) increased separation between film and substrate due to insertion of the graphene and (2) attenuation due to free carrier screening. 
We start from an empirical Morse interatomic potential \cite{morse1929diatomic}, which includes covalent, vdW, and electrostatic interactions \cite{mirzanejad2025derivation}, to describe the substrate. This simple two body potential captures the essential physics and employs easy to understand parameters that are well benchmarked by experiments, density functional theory, and molecular dynamics simulations \cite{keyes1975bonding}, before the inclusion of graphene. The Morse potential has the form
\begin{equation}
    \phi_{sub}(z) = D \left[ e^{-2a(z-z_0)} - 2e^{-a(z-z_0)} \right]
\end{equation}
where the first term describes the short-range Pauli repulsion and the second term is the long range attraction. Here $D$ is the bond strength, $1/a$ is the width of the potential well, and $z_0$ is the equilibrium bond distance.
The potential for a bare GaAs substrate without a remote spacer layer is shown in Fig. \ref{fig:screening}(a) (black curve). This potential describes the pair bonding interaction between an As atom and a Ga atom, and thus is most appropriate for homoepitaxy of a GaAs film on a GaAs substrate (or to a limited extent, heteroepitaxy with Ga-As interface termination, e.g. InAs/GaSb (001)). We use parameters $D = 2.1$ eV and $a = 1.52$ \AA$^{-1}$, based on optical spectroscopy measurements of the bond strength and oscillation frequency of GaAs \cite{lemire1990spectroscopy}, and refined by ab initio calculations in a many-body form to reproduce the bulk properties \cite{albe,Fichthorn}. We let $ z_0 = 3$ \AA, corresponding to a generalized layer spacing between atomic planes.

% morse_screen(D,   z0, z1,     a0,             a1,     density, thick)
% morse_screen(2.1, 3,  3*1.65, 1.52,   1.52*1.6,       1e20, 3)
We next model the insertion of an insulating 2D material spacer at distance $z=z_0$ from the substrate, in the absence of screening. This is analogous to inserting monolayer hBN, to produce a new minimum at increased distance from the substrate. Since the primary out of plane interactions for 2D materials are van der Waals, we model the 2D material with a Lennard-Jones potential 
\begin{equation}
    \phi_{2D}(z-z_0) = 4\epsilon \left[ \left(\frac{\sigma}{z-z_0}\right)^{12} - \left(\frac{\sigma}{z-z_0}\right)^6 \right]
\end{equation}
where we let $\epsilon = 5$ meV, $\sigma = 3.2 \times 2^{-1/6}$ \AA, and $z_0$ is the equilibrium position of the new 2D material spacer. The total potential $ \phi_{sub}(z) + \phi_{2D}(z-z_0)$, in the absence of screening, is shown in the blue curve. The inert spacer displays a new minimum of $-60$ meV, which is only $3\%$ of the Ga-As bond energy. The new equilibrium film-substrate separation of $z=5.7$ \AA\ is within the 5-6 \AA\ separation typically observed by TEM of GaAs/graphene/GaAs \cite{kim2021role}. Our insulating spacer calculation is consistent with previous slab DFT calculations that use charge density as a proxy for the remote potential permeation: for (001)-oriented GaAs slabs separated by a 7 \AA\ vacuum gap (no graphene for screening), the computed charge density in the middle of the gap was $\sim 4\%$ of the charge density in the interior of the material \cite{kim2017remote}. 

We now include free carrier screening from metallic graphene. This screening is typically ignored by slab DFT calculations that either do not include or try to subtract out the contributions from graphene \cite{kim2017remote, kong2018polarity,dai2022highly}. We model the graphene as a homogeneous electron gas with Thomas-Fermi wavenumber $k_{TF}^2 = (3^{1/3}me^2 n^{1/3})/(\epsilon_0 \hbar^2 \pi^{4/3})$
% \begin{equation}
%     k_{TF}^2 = \frac{3^{1/3}me^2 n^{1/3}}{\epsilon_0 \hbar^2 \pi^{4/3}}
% \end{equation}
where $n$ is the carrier density, $m_0$ is the electron rest mass, and $\hbar$ is the reduced Planck constant.
The screened transmission coefficient through graphene is 
\begin{equation}
    T_s = exp(-k_{TF} N \Delta z_{ML})
\end{equation} % rewrite as n \Delta z, where n is number of graphene layers?
where $\Delta z_{ML}= 3$ \AA\ is the effective thickness of graphene and $N$ is the number of graphene layers. Figure \ref{fig:screening}(c) plots $T_s$ through monolayer and bilayer graphene, as a function of typical room temperature graphene carrier densities $n_{2D} = 10^{11} - 10^{14}$ cm$^{-2}$ ($n_{3D} = n_{2D}/\Delta z_{ML}$). We find that the screening is highly tunable with carrier density, and there is a significant screening-induced attenuation going from monolayer graphene to bilayer graphene (in addition to the attenuation from increased layer separation). 

For $z$ greater than the thickness of graphene, the total screened potential is 
\begin{equation}
    \phi_{total}(z) = T_s \phi_{sub}(z) + \phi_{2D}(z-z_0)
\end{equation}
where the first term $\phi_{remote} = T_s \phi_{sub}$ is the screened remote potential of the substrate and the second term is the Lennard-Jones (vdW) potential of graphene.
Remote epitaxy is expected if $T_s \phi_{sub} > \phi_{2D}$, i.e. if the screened substrate potential dominates over the graphene potential. This would produce an epitaxial film aligned to the substrate. Van der Waals epitaxy is expected if $\phi_{2D}>T_s \phi_{sub}$, resulting in a film that grow in the lowest surface energy orientation, with a weak driving force for in-plane alignment.

Fig. \ref{fig:screening}(b) plots $\phi_{total}(z)$ for a range of typical graphene carrier densities. Here our model provides two testable predictions. First, the equilibrium film-substrate separation (minimum of $\phi_{total}$) shifts to larger $z$ with increased carrier density (screening). Thus the separation can serve as a proxy for the remote interaction strength. For very large graphene carrier density, or as $T_s$ approaches 0, $\phi_{total}$ converges to the Lennard-Jones potential of graphene $\phi_{2D}$. Electrostatic gating of the graphene could be used to test this prediction.

Second, the magnitude of $\phi_{total}$ is significantly attenuated by screening. For a carrier density of $n_{2D} = 3\times 10^{12}$ cm$^{-2}$, the magnitude of the $\phi_{total}$ at the new minimum of $z=6$ \AA\ is 15 meV. This magnitude is $140 \times$ smaller than the GaAs covalent bond strength ($D=2.1$ eV), smaller than typical surface diffusion barrier heights (100 meV), and comparable to the meV scale for van der Waals interactions (Fig. \ref{fig:screening}(c)). Note that the template for epitaxy is given by lateral fluctuations in the potential $\Delta \phi(x,y)$ rather than the full magnitude of the total potential $|\phi_{total}|$. This suggests that the lateral fluctuations $\Delta \phi(x,y)$ are smaller than 15 meV. 

Our calculations reveal that once screening is included, the remote potential of GaAs through monolayer graphene is not the dominant interaction. Rather, $|\phi_{remote}|$ has similar magnitude as the meV scale van der Waals interaction of graphene itself, and hence the graphene potential cannot be ignored in remote epitaxy. Since this screening depends exponentially on the thickness of the metallic layer $N\Delta z_{ML}$, the prospect for ``remote'' epitaxy through bilayer or even trilayer graphene is vanishingly weak, in contrast with previous experimental reports of remote epitaxy through several layers of graphene \cite{kong2018polarity}. Note that most experiments cannot rule our pinhole nucleation as the mechanism rather than remote interactions, since contaminants and oxides at the transferred graphene/substrate interfaces typically create pinholes in the graphene when heated to film growth temperatures \cite{manzo2022pinhole}.

At first glance this may seem inconsistent with previous DFT slab calculations of the permeated potential, which suggested 15 meV fluctuations through graphene on GaAs substrates \cite{kong2018polarity}. However, we emphasize that (1) the previous calculation was for the electrostatic potential \cite{kong2018polarity}, which is not the bonding potential, 
(2) the calculations in Ref. \cite{kong2018polarity} employed selective cropping and replication and thus are not a reliable representation of the potential fluctuations \cite{laduca2025transparent}, and (3) in many calculations the screening effects from graphene and the lattice potential of graphene itself are not explicitly included.

\begin{table}[]
\begin{tabular}{l|c|c|c|c|c}
 & $\Delta \chi$ & $D$ (eV) & $a$ (\AA$^{-1}$) & $|\phi_a/D|$ & $|\phi_{remote}|$ (meV) \\
\hline
ZnO      & 1.79  & 3.60     & 1.82     & $8.5 \times 10^{-3}$   & 5.3 \\
GaN      & 1.63  & 2.45     & 1.97     & $5.5 \times 10^{-3}$   & 2.3 \\
GaAs     & 0.37  & 2.10     & 1.52     & $2.1 \times 10^{-2}$   & 7.6 \\
SiC      & 0.65  & 4.36     & 1.70     & $1.2 \times 10^{-2}$   & 9.3 \\
Si       & 0     & 3.24     & 1.48     & $2.4 \times 10^{-2}$   & 13  \\
Ge       & 0     & 1.63     & 1.50     & $2.2 \times 10^{-2}$   & 6.2  \\
Cu       & 0     & 0.34     & 2.87     & $3.7 \times 10^{-4}$   & 0.022 \\
\hline
\end{tabular}
\caption{Electronegativity difference, Morse parameters, distance induced attenuation, and remote potential strength through monolayer graphene for several commonly used substrates \cite{erhart2006analytic, nord2003modelling, albe, Fichthorn, erhart2005analytical, van2011study, girifalco1959application}.}
\end{table}

We now apply our model to other commonly used semiconductor and oxide substrates. Table 1 summarizes the Morse potential parameters for 
ZnO \cite{erhart2006analytic}, GaN \cite{nord2003modelling}, GaAs \cite{albe,Fichthorn}, SiC \cite{erhart2005analytical}, Si \cite{erhart2005analytical}, Ge \cite{van2011study}, and Cu \cite{girifalco1959application}.  
We first use these parameters to estimate the attenuation from the separation induced by the offset of an inert 2D materials spacer, compared to the attenuation from screening. For simplicity, we use the fact that for large $z$ the attractive term of the Morse potential (Eqn 1) dominates. We thus use $\phi_{sub}(z) /D \approx \phi_{a}(z)/D = -2 \, e^{-a(z-z_0)}$. In Table 1 we compute $\phi_{a}(z)/D$ at $z=6$ \AA. We find that the separation induced attenuation for these substrates of $10^{-4}$ to $10^{-2}$ is slightly more significant than the screening induced $T_s = 10^{-2} - 10^{-1}$ for monolayer graphene at typical carrier densities $n_{2D} = 10^{11} - 10^{14}$ cm$^{-2}$ (Fig. \ref{fig:screening}(b)).

We next test the hypothesis from Ref. \cite{kong2018polarity} that polar ionic bonding controls the strength of the remote potential. For a fixed graphene carrier density of  $n_{2D}=3\times 10^{12}$ cm$^{-2}$, Table 1 summarizes the remote potential under the approximation 
\begin{equation}
    |\phi_{remote}| \approx 2D \,exp\left[-k_{TF} N\Delta z_{ML}  -a(z-z_0)\right].
\end{equation}

\begin{figure}
    \centering
    \includegraphics[width=1\linewidth]{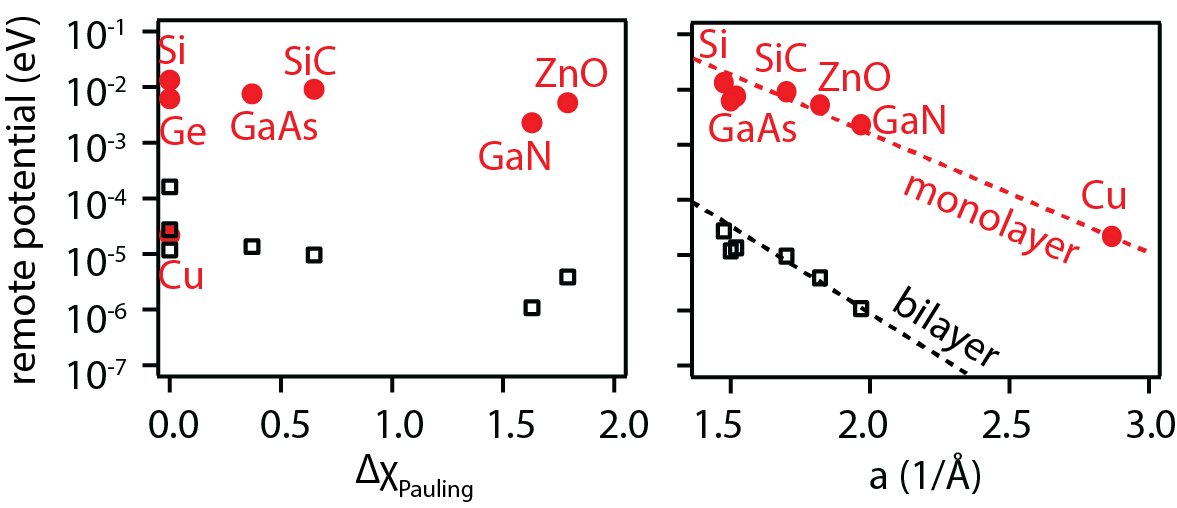}
    \caption{Estimated remote potential through monolayer and bilayer graphene for commonly used substrates, at fixed distance of $z=6$ \AA\ and 9 \AA, respectively.}
    \label{fig:materials}
\end{figure}

We plot this magnitude versus the Pauling electronegativity difference $\chi$ in Fig. \ref{fig:materials}, where we let $z= z_0 + N\Delta z_{ML}$. We find no clear trend of $|\phi_{remote}|$ with $\chi$. On the other hand, a clear trend emerges with the Morse potential width $1/a$, where more spatially extended bond potentials (small $a$) lead to a larger remote potential at a distance of 6 \AA\ from the substrate. This trend is clearly seen by the shape of the potential (Eqn 1), where the potential varies exponentially with $a$ but only linearly on the bond strength $D$. 

Interestingly, our model suggests that monoatomic Si should have the strongest remote potential among the substrates considered. 
This lies in apparent contradiction to the experiments in Ref. \cite{kong2018polarity} reported that growth of Si on graphene/Si (001) resulting in a polycrystalline film, yet growth on substrates with more polar bonding enabled epitaxial films in more ionic LiF substrates that were covered with up to three layers of graphene. However, the impact of key synthesis parameters including substrate temperature, fluxes, type of precursor (e.g., atom, molecular cluster, metalorganic), and graphene surface preparation remain unclear. We also note that in this case, carbon in the graphene may bond with the silicon substrate, altering the graphene potential from the straightforward screening assumption of our model.

Our model calls for a re-examination of Si and other substrates with small $a$, coupled with a systematic understanding of how synthesis conditions and graphene surface preparation affect the resulting films and their crystalline orientations. The model suggests that the best substrates for remote epitaxy are those with spatially extended bond potentials (or spatially extended orbitals), rather than substrates with more ionic bonding character. Our calculations also suggest that tuning the screening via gating the graphene may be a powerful route to tune the strength of the remote potential.

We caution that our model should only be considered an order of magnitude estimate for the remote potential. We treat graphene as an idealized homogeneous electron gas, and only consider pair-wise atom-atom interactions between film and substrate. Our model does not consider the directional character of bonding, multi-atom interactions, or bonding between the film and graphene or graphene and the substrate. It also does not consider surface reconstructions. For lattice mismatched heteroepitaxy, it does not consider strain or misfit dislocation energies. Still, it identifies an essential role for the decay with distance and for screening. These essential parameters place significant constraints on the magnitude of the remote bonding potential that are overlooked in the current atomistic calculations. 
More accurate state-of-the-art bond potentials, e.g. Tersoff potentials used in molecular dynamics simulations, could be applied to our framework of Eq. 4 to enhance our the model's predictive power. Note, however, that more complex yet accurate potentials come at the cost of losing simple physically interpretable parameters. Additionally, future work could move Eq. 4 from a 1D model of bonding to a 2D model which could allow for the incorporation of additional physics such as reconstructions.

Our calculations also highlight the need for direct measurements of the remote potential. Current experiments rely on the final outcomes of growth to infer the strength of the remote potential through graphene. However, such experiments convolve the impact of the remote potential with defects and variable growth kinetics, thus obscuring an understanding of the growth mechanisms. Our model, which contains physically understandable parameters of potential width $1/a$, potential depth $D$, and screening $T_s$ in an analytical form, facilitates comparisons with direct measurements to benchmark our theory. We propose frequency-shift atomic force microscopy \cite{ke1999quantity, hoffmann2001direct, giessibl2001imaging, allain2017color} and He atom scattering \cite{boato1979bound} as methods to directly measure the covalent and vdW contributions, to complement the Kelvin-probe force microscopy \cite{filleter2008local} and photoemission/low energy electron microscopy (PEEM/LEEM) \cite{hibino2009dependence} that have mapped the electrostatic potential (workfunction) above graphene-covered substrates. These measurements could be performed as a function of gate-induced graphene screening to test our predictions of Eq 4. DFT calculations could also be used to computationally examine these remote potentials, provided they fully include any substrate-graphene reconstruction, the spacer layer of graphene, and at least one atom of the film layer to measure full chemical bonding and not just electrostatic potential.

\section{Acknowledgments}

Model development by JKK was primarily supported by the U.S. Department of Energy, Office of Science, Basic Energy Sciences, under award no. DE-SC0023958. Comparison of different substrates was supported by the Sandia University Partnerships Network (SUPN) program of Sandia National Laboratories.

QTC is supported by the by the Laboratory Directed Research and Development (LDRD) program at Sandia National Laboratories under project 233271.
This work was performed, in part, at the Center for Integrated Nanotechnologies, an Office of Science User Facility operated for the U.S. Department of Energy (DOE) Office of Science.
Sandia National Laboratories is a multi-mission laboratory managed and operated by National Technology \& Engineering Solutions of Sandia, LLC (NTESS), a wholly owned subsidiary of Honeywell International Inc., for the U.S. Department of Energy’s National Nuclear Security Administration (DOE/NNSA) under contract DE-NA0003525. This written work is authored by an employee of NTESS. The employee, not NTESS, owns the right, title and interest in and to the written work and is responsible for its contents. Any subjective views or opinions that might be expressed in the written work do not necessarily represent the views of the U.S. Government. The publisher acknowledges that the U.S. Government retains a non-exclusive, paid-up, irrevocable, world-wide license to publish or reproduce the published form of this written work or allow others to do so, for U.S. Government purposes. The DOE will provide public access to results of federally sponsored research in accordance with the DOE Public Access Plan.

\bibliographystyle{apsrev}
\bibliography{ref}

\end{document}